\documentstyle[prl,aps,axodraw,floats]{revtex}
\def\bi{\bibitem}

\newcommand{\para}{_\parallel}
\newcommand{\pr}{_\perp}

\begin{document}
\draft
\twocolumn[
\hsize\textwidth\columnwidth\hsize\csname@twocolumnfalse\endcsname
\title{Neutrino Photon Interaction in a Magnetized Medium II}
\begin{flushright}
IMSc/2002/07/21\\
SINP/TNP/02-28
\end{flushright}
\bigskip
\author{Kaushik Bhattacharya$^1$,~~Avijit K.~Ganguly$^2.$}
\address{$^1$ Saha Institute of Nuclear Physics, 1/AF, Bidhan-Nagar, Calcutta
700064, India \\
$^2$ Institute of Mathematical Sciences, Chennai.\\
e-mail :kaushikb@theory.saha.ernet.in,avijit@imsc.ernet.in, }
\maketitle
\begin{abstract}
In the presence of a thermal medium or an external electro-magnetic field,
neutrinos can interact with photon, mediated by the corresponding charged 
leptons (real or virtual). The effect of a medium or an electromagnetic 
field is two-fold - to induce
an effective $\nu \gamma$ vertex and to modify the dispersion
relations of all the particles involved to render the processes
kinematically viable. It has already been noted that in a medium
neutrinos acquire an effective charge, which in the standard model
of electroweak interaction comes from the vector type vertex of
weak interaction. On the other hand in a magnetized plasma, the
axial vector part also start contributing to the effective charge
of a neutrino. This contribution corresponding to the axial vector
part in the interaction Lagrangian  is denoted as the axial polarisation tensor. 
In an earlier paper we explicitly calculated the form of the axial
polarisation tensor  to all odd orders in external magnetic field.
In this note we complete that investigation by computing the same,
to all even orders in external magnetic field.  We further show its 
gauge invarience properties. Finally we infer upon the zero external 
momentum limit of this axial polarisation tensor. 
\end{abstract}
\pacs{PACS numbers::~11.10Wx, 12.20-m, 13.15+g, 97.10Ld}
\narrowtext
\bigskip
]
\section{Introduction}
\label {intr}
Neutrino mediated processes are of great importance in  
cosmology and astrophysics{\cite{raff,weinberg}}. 
It is worth mentioning at this stage that various interesting
possibilities have been looked into in the context  of cosmology
e.g, large scale structure formation in the universe, to name one  of
the few~\cite{palone}. In this note we would rather consider the astrophysical part of it.

 Because of the effective
neutrino photon interaction in a medium, it is possible that the
neutrinos might dump a fraction of their energy inside the star during
stellar evolution. For instance in a type II supernovae collapse
\cite{colp} neutrinos  produced deep inside the proto-neutron star
surge out carrying an effective energy $\sim 10^{52}$ erg/s. It is
conjectured that the neutrinos deposit some fraction of its energy
during the explosion through different kinds of neutrino electromagnetic
interactions, e.g,  $\gamma \nu \to \gamma \nu $, $\nu \to \nu \gamma$, 
$\nu \bar{\nu} \to e^{+} e^{-}$,
 to name a few.  It is important to note here, that all these 
processes are of order $G^2_F$.  However the amount of energy dumped
by these mechanisms to the mantle of the proto-neutron star  seem to 
be barely sufficient to blow the outer part of the same.
However, it is  important to note that,  inside a star,  a
nonzero magnetic field would always be present, which
in most of the studies performed so far, has not  been properly
taken into account.   Our objective here is to take into account
the presence of a strong magnetic field and estimate the corrections
coming there off. 
  
It is usually conjectured, taking into account the conservation of
surface magnetic field of a proto-neutron star, that during a
supernova collapse  the magnetic field strength in some regions inside
the nascent star  can reach upto ${\cal B}\sim \frac{ m^2}{e} $ or
more. Here $m$ denotes the mass of electron. [Henceforth we would refer to field
strengths of this magnitude as critical field strength ${\cal
B}_c$.]  This conjecture makes it worthwhile to investigate the
role of magnetic field in effective neutrino photon vertex.

Neutrinos do not couple with photons in the
tree level in the standard model of particle physics, and this
coupling can only take place at a loop level, mediated by the fermions
and gauge bosons. This coupling can give birth to off-shell photons
only,  since for on-shell particles, the processes like $\nu \to \nu
\gamma$ and $\gamma \to \nu \bar{\nu}$ are restrained
kinematically. Only in presence of a medium can all the particles be
onshell as there the dispersion relation of the photon changes, 
giving the much required phase space for the reactions.   
Intuitively when a neutrino moves inside a thermal medium composed of electrons
and positrons, they interact with these background particles. The
background electrons and positrons themselves have interaction with
the electromagnetic fields, and this fact gives rise to an effective
coupling of the neutrinos to the photons. Under these circumstance's the
neutrinos may acquire an ``effective electric charge'' through
which they interact with the ambient  plasma.

In this paper we concentrate upon the effective neutrino photon 
interaction vertex coming from the axial vector part of the interaction.
From there we estimate the
effective charge of the neutrino inside a magnetised medium. 
We name the axial contribution in the effective neutrino photon Lagrangian
as the axial polarisation tensor $\Pi^5_{\mu \nu}$, which we will clearly
define in the next section. We discuss the physical situations where the axial polarisation
tensor arises, then show how it affects the physical processes. The
effective charge of the neutrino has been calculated previously by
many authors ~\cite{pal1,orae,alth}.  In this regard we also
comment upon the contribution of this quantity on the effective charge
of the neutrinos inside a magnetised plasma as well as other physical
observables.

The plan of the paper is as follows.  
We start with section \ref{form} that deals with the
formalism through which the physical importance of $\Pi^5_{\mu \nu}$
is appreciated. In the next section \ref{gd}
general form factor analysis of ``axial polarisation tensor'' 
on the basis of symmetry arguments is provided.
In section \ref{capt} we show the fermion propagator in a
magnetised medium, and using the same explicitly write down
$\Pi^5_{\mu \nu}$ in the rest frame of the medium.  In the section after,
i.e section \ref{gaugeinvariance}, we
show the formal proof of gauge invariance 
of $\Pi^5_{\mu\nu}$.
 In Section \ref{efftcharge}  we
calculate the effective electric charge from the expression of the
axial polarisation tensor. In section \ref{concl} we discuss our 
results and conclude 
by touching upon the physical relevance of our work.
\section{Formalism}
\label{form}
%
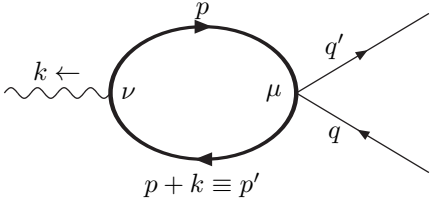
\begin{figure}
\begin{center}
\begin{picture}(150,40)(0,-35)
\Photon(40,0)(0,0){2}{4}
\Text(20,5)[b]{$k\leftarrow$}
\Text(75,30)[b]{$p$}
\Text(125,-17)[b]{$q$}
\Text(125,15)[b]{$q'$}
\Text(47,0)[]{$\nu$}
\Text(102,0)[]{$\mu$}
\Text(75,-30)[t]{$p+k\equiv p'$}
\ArrowLine(110,0)(160,30)
\ArrowLine(160,-30)(110,0)
\SetWidth{1.2}
\Oval(75,0)(25,35)(0)
\ArrowLine(74,25)(76,25)
\ArrowLine(76,-25)(74,-25)
\end{picture}
\end{center} 
\caption[]{One-loop diagram for the effective electromagnetic vertex of the neutrino in the limit of infinitely
heavy W and Z masses.}\label{f:cher}
\end{figure}
In this work we consider neutrino momenta that are small compared to
the masses of the W and Z bosons. We can, therefore, neglect the
momentum dependence in the W and Z propagators, which is justified if
we are performing a calculation to the leading order in the Fermi
constant, $G_F$. In this limit four-fermion interaction is given by
the following effective Lagrangian:
\begin{eqnarray}
{\cal L}_{\rm eff} = -\frac{1}{\sqrt{2}} G_F {\overline \nu}
\gamma^{\mu} (1 - \gamma_5) \nu {\overline l_\nu} \gamma_{\mu} (g_{\rm
V} + g_{\rm A} \gamma_5) l_\nu \,,
\end{eqnarray}
where, $\nu$ and $l_\nu$ are the neutrino and the corresponding lepton field respectively. For electron neutrinos,
\begin{eqnarray}
g_{\rm V} &=& 1 - (1 - 4 \sin^2 \theta_{\rm W})/2, \\
g_{\rm A} &=& -1 + 1/2;
\end{eqnarray}
where the first terms in $g_{\rm V}$ and $g_{\rm A}$ are the
contributions from the W exchange diagram and the second one from the
Z exchange diagram.

With this interaction Lagrangian we can write down the matrix element
for the Cherenkov amplitude as,
\begin{eqnarray}
M = - \frac{G_F}{\sqrt{2} e} Z \epsilon^{\nu} \bar{\nu}\gamma^{\mu} (1
- \gamma_5) \nu (g_V \Pi_{\mu \nu} + g_A \Pi^5_{\mu \nu})
\label{cheren}
\end{eqnarray}
where $\epsilon^{\nu}$ is the photon polarisation tensor, and $Z$ is
the wavefunction renormalisation factor inside a medium. The term
$\Pi_{\mu \nu}$ is defined as
\begin{eqnarray}
i\Pi_{\mu \nu}=(-i e)^2(-1) \int {{d^4 p}\over {(2\pi)^4}}\mbox{Tr}\left
[ \gamma_\mu iS(p) \gamma_\nu iS(p')\right]
\label{npimunu}
\end{eqnarray}
which looks exactly like the photon polarisation tensor, but dosent
have the same interpretation here. The momentum labels of the
prapagators can be understood from fig.[\ref{f:cher}].   Henceforth we would call it 
the polarisation tensor.  $\Pi^5_{\mu \nu}$ is defined as 
\begin{eqnarray}
i\Pi^5_{\mu \nu}=(-i e)^2(-1) \int {{d^4 p}\over {(2\pi)^4}}\mbox{Tr}\left
[ \gamma_\mu \gamma_5 iS(p) \gamma_\nu iS(p')\right]
\label{npimunu5}
\end{eqnarray}
which we call the axial polarisation tensor. Both the polarisation tensor
and the axial polarisation tensor are obtained by
calculating the Feynman diagram given in
fig.[\ref{f:cher}]. 

The off-shell electromagnetic vertex function $\Gamma_{\nu}$ is
defined in such a way that, for on-shell neutrinos, the $\nu \nu
\gamma$ amplitude is given by:
\begin{eqnarray}
{\cal M} = - i \bar{u}(q') \Gamma_{\nu} u(q) A^{\nu}(k),
\label{chargedef}
\end{eqnarray}
where, $k$ is the photon momentum. Here, $u(q)$ is the 
the neutrino spinor and $A^{\nu}$ stands for the electromagnetic vector potential. In general $\Gamma_{\nu}$ 
would depend on $k$  and the characteristics of the medium. With our
effective Lagrangian $\Gamma_{\nu}$ is given by
\begin{eqnarray}
\Gamma_{\nu} = - \frac{1}{\sqrt{2}e} G_F \gamma^{\mu} (1 - \gamma_5) \,(g_{\rm V} \Pi_{\mu \nu} + g_{\rm A} \Pi_{\mu \nu}^5) \,,
\end{eqnarray}

The effective charge of the neutrinos is defined in terms of the
vertex function by the following relation~\cite{pal1}:
\begin{eqnarray}
e_{\rm eff} = {1\over{2 q_0}} \, \bar{u}(q) \, \Gamma_0(k_0=0, {\bf k}
\rightarrow 0) \, u(q) \,.
\label{chargedef1}
\end{eqnarray}
For massless Weyl spinors this definition can be rendered into the form:
\begin{eqnarray}
e_{\rm eff} = {1\over{2 q_0}} \, \mbox{Tr} \left[\Gamma_0(k_0=0, {\bf
k}\rightarrow 0) \, (1+\lambda \gamma^5) \, \rlap/q \right]
\label{nec1}
\end{eqnarray}
where $\lambda = \pm 1$ is the helicity of the spinors.

While discussing about $\Pi^5_{\mu \nu}$ it should be remembered that for
the electromagnetic vertex, we  have the current conservation relation,
\begin{eqnarray}
k^{\nu} \Pi^5_{\mu \nu} = 0
\label{ngi}
\end{eqnarray}
which is the gauge invariance condition.

In order to calculate the Cherenkov amplitude or
 the effective charge of the neutrinos inside a medium,
we have to calculate $\Pi^5_{\mu \nu}$. The formalism so discussed is
a general one and we extend the calculations prviously done based
upon this formalism to the case where we have a constant background
magnetic field in addition to a thermal medium. In doing so
 we give the full
expression of $\Pi^5_{\mu \nu}$ in a magnetised medium and
explicitly show its gauge invariance. We also comment on the
effective charge contribution from the axial polarisation part.  

Discussing about the effective charge of the neutrinos in a
medium,  the way we have done, it should be mentioned that although it is
interesting to find it theoretically, it is not the ``charge'' with
which the neutrinos couple with a magnetic field. From the
definition of the electromagnetic vertex as given in
Eq.(\ref{chargedef}) and the definition of charge in
Eq.(\ref{chargedef1}) it is clear that we are interested to find
the coupling of the photon field with $\bar{u}\Gamma^0 u$ and not
$\bar{u}\Gamma^i u$. The magnetic 
interaction will come from the term $\bar{u}(q)\Gamma^i u(q) A_i$, but
we will see at 
the end of the calculation that no $\Gamma^i$ exists in the limit $k_0
\to 0, \vec{k} \to 0 $, whereby we cant say of any possible 
interaction of the neutrinos with the external static uniform magnetic field.

\section{General Analysis}
\label{gd}
We start this section with a discussion on the possible tensor
structure and form factor analysis of $\Pi^{5}_{\mu\nu}(k)$, based on
the symmetry of the interaction.
To begin with we note that, $\Pi^{5}_{\mu\nu}(k)$ in vacuum should vanish.
This follows from the following arguments. In vacuum the available vectors
and tensors at hand  are  the following,
\begin{eqnarray}
k_{\mu},~g_{\mu\nu} \mbox{~and~} \epsilon_{\mu\nu\lambda\sigma}.
\label{vt}
\end{eqnarray}

The two point axial-vector correlation function
$\Pi^{5}_{\mu\nu}$ can be expanded in a basis, constructed out of
tensors $g_{\mu\nu}$, $\epsilon_{\mu\nu\lambda\sigma}$, and
vector $k_{\lambda}$.  Given the parity structure of the theory it is
impossible to construct a tensor of rank two using $g_{\mu\nu}$ and
$k_{\mu}$ . 
So the only available tensor (with the right parity structure)
  we have at hand is  $\epsilon_{\mu\nu\lambda\sigma}$.
The  other vector needed to make it a tensor of rank two is
 $k_{\lambda}$.  As we contract
$\epsilon_{\mu\nu\lambda\sigma}$ with $k_{\lambda}$, $k_{\sigma}$,
since $\epsilon_{\mu\nu\lambda\sigma}$ is completely antisymmetric
tensor of rank four,  the corresponding term vanishes.

On the other hand, in a medium, we have an additional vector
 $u^{\mu}$, i.e  the velocity of the centre of mass of the medium.
Therefore  the polarisation tensor can be expanded in terms of form
factors along with the new tensors constructed out of $u^{\mu}$
and the ones we already had in absence of a medium as,
\begin{eqnarray}
\Pi_{\mu\nu}(k)=\Pi_T\,T_{\mu\nu} + \Pi_L\,L_{\mu\nu}. 
\end{eqnarray}
Here
\begin{eqnarray}
T_{\mu\nu}&=& {\widetilde g}_{\mu\nu} - L_{\mu\nu}\\
L_{\mu\nu}&=&\frac{{\widetilde u}_{\mu}{\widetilde
u}_{\nu}}{{\widetilde u}^2}\\
\label{t}
\end{eqnarray}
with
\begin{eqnarray}
{\widetilde g}_{\mu\nu}&=&g_{\mu\nu} - \frac{k_{\mu}k_{\nu}}{k^2}\\
{\widetilde u}_{\mu}&=&{\widetilde g}_{\mu\rho} u^{\rho}\\
\end{eqnarray}
In the rest frame of the medium  the four velocity is given by $u^{\mu}=(1,0,0,0)$.
It is easy to see that the longitudinal projector $L_{\mu \nu}$ is not
zero in the limit $k_0=0,\vec{k}\rightarrow 0$ and $\Pi_L$ is also not
zero in the above mentioned limit. This fact is responsible for giving
nonzero contribution to the effective charge of
neutrino. 

As has already been mentioned, that  in a medium, we have another extra four vector
 $u^{\mu}$ and hence  it is possible to construct the axial polarisation tensor
 of rank two, out of $\varepsilon_{\mu \nu
\alpha \beta}$, $u_{\mu}$, $k_{\mu}$, i.e, $\varepsilon_{\mu \nu
\alpha \beta}u^{\alpha}k^{\beta}$, that would verify the Ward identity
for the two point function. An explicit calculation of
$\Pi^{5}_{\mu\nu}(k)$ verifies the tensor structure  of it as predicted
here. It is worth noting that this contributes to the Cherenkov  
amplitude, but not to the effective electric charge of the neutrinos 
since for charge
calculation we have to put the index $\nu = 0$. In the rest frame
only $u^{0}$ exists, that forces the totally antisymmetric tensor 
to vanish. 

In a constant background magnetic field in addition to the ones mentioned in
Eq. (\ref{vt}) one has the freedom of having other   
extra vectors and tensors (to first order in field strength), such as 
\begin{eqnarray}
F_{\mu\nu},~{\widetilde{F}_{\mu\nu}}
\end{eqnarray}
along with
\begin{eqnarray} 
e^{(1)}_\mu= k^{\lambda}F_{\lambda\mu},~~~~~
e^{(2)}_\mu = k^\lambda{\widetilde{F}_{\lambda\mu}}.
\label{vtm}
\end{eqnarray}
Explicit evaluation of the axial polarisation tensor, in a constant background magnetic
field is (however the metric used by
the authors in refernces mentioned  is different from that of us) {\cite{hari,iorf}},
\begin{eqnarray}
& &\Pi^{5}_{\mu\nu}(k) = \frac{e^3}{(4\pi)^2 m^2}\left[-C\para
k_{\nu\para} ({\widetilde F}k)_{\mu}\right.\nonumber\\
& &\left.\hskip 1cm +\,\, C\pr\left\{k_{\nu\pr}
(k{\widetilde F})_{\mu} + k_{\mu\pr}(k{\widetilde F})_{\nu} -
k^2\pr{\widetilde F}_{\mu \nu}  \right\}\right]
\label{harid}
\end{eqnarray}
where ${\widetilde F}^{\mu \nu} = \frac{1}{2}{\varepsilon^{\mu \nu
\rho \sigma}} F_{\rho \sigma}$, $F_{1 2}=-F_{2 1}= {\cal B}$ and 
$(k{\widetilde F})_{\nu}={\widetilde F}^{\mu \nu} k_{\mu} .$
According to the notation used in Eq.(\ref{harid}),  $k\para =
(k_0,0,0,k_3)$ and $k\pr=(0,k_x,k_y,0)$. Lastly
$C\para$ and $C\pr$ are functions of ${\cal B}, k^2\para,k^2\pr$. 
It is easy to note that in consonance with the general parity structure
of the theory the basis tensors for this case are ${\widetilde{F}_{\mu\nu}}$,
$e^{(2)}_{\mu}k_{\nu\pr}$ and $e^{(2)}_{\nu}k_{\mu\pr}$  . 

From the above expression we can see that the axial polarisation tensor in
a background magnetic field does not survive when the momentum of the external
photon vanishes, and as a result there cannot be any effective
electric charge of the neutrinos in a constant background magnetic field.
Actually this formal statement could have been spoilt by the 
presence of possible infrared divergence in the loop; i.e
to say in $C\para$ and $C\pr$. Since the particle inside 
the loop is massive so there is no scope of having infrared 
divergence, hence it doesn't contribute to neutrino effective charge.

In the presence of an external magnetic field  (to even and odd order in field strength) 
plus medium, we have other vectors available. The ones important for
our purpose,  are the following,
\begin{eqnarray}
e^{(3)}_{\mu} &=& F_{\mu\alpha}F^{\alpha\beta}k_{\beta} \nonumber \\
e^{(4)}_{\mu} &=&\epsilon_{\mu\nu\lambda\sigma}F^{\lambda\sigma}u^{\nu}.
\label{vtmb}
\end{eqnarray}
It is worth noting that, in a background magnetic field pointed in the z direction,
the possible vector with highest power of external magnetic field that
can occur in the axial polarisation tensor is  $e^{(3)}_{\mu}$. Other
constructions with higher powers 
of field strength tensor would simply be a linear combination of the same
and/or $F^{\mu\nu}k_{\nu}$. 
The structure of axial polarisation tensor for odd powers in magnetic field
 , had already been analysed in Ref.{\cite{kas}}, so we would directly comment
on the same with even powers of external magnetic field ${\cal B}$.
Since the underlying theory is CP conserving, therefore, the possibilities
of decomposing $\Pi^{5}_{\mu\nu}$ in terms of the basis vectors are,
\begin{eqnarray}
\Pi^{5(O({\cal B}^{2}))}_{\mu\nu} = F_{1} \epsilon_{\mu\nu\lambda\sigma}
u^{\lambda}k^{\sigma}+ F_{2}\epsilon_{\mu\nu\alpha\beta}k^{\alpha}e^{(3)\beta},  
\label{pioo}
\end{eqnarray}
where $F_1$ and $F_2$, i.e. the form factors, are functions of Lorentz 
scalars, constructed out of all the vectors or tensors we have at our 
disposal as well other parameters like temperature and chemical potential.
 The  first term on the right hand side of Eq.(\ref{pioo}) have already been
discussed in \cite{pal1}, with the exception that the function $F_{1}$
 now is a  even function of external magnetic field ${\cal B}$; on the 
other hand the appearence of the second term is new. One can also observe 
that, in keeping with CP invariance of the theory ( i.e background along 
 with the interaction),  both the functions $F_1$ and  $F_2$ should be 
odd functions of chemical potential. However
this would become clear from  Eq.(\ref{evenpart}) of section \ref{gaugeinvariance}.
 
\section{One  loop calculation of the  axial polarisation tensor}
\label{capt}
Since we investigate the case with a background magnetic field,  without any loss of
generality it can be
taken to be in the $z$-direction. We denote the magnitude of this field by $\cal
B$. Ignoring  first the presence of the medium, the electron
propagator in such a field can be written down following Schwinger's
approach~\cite{schw,tsai,ditt}:
\begin{eqnarray}
i S_B^V(p) = \int_0^\infty ds \, e^{\Phi(p,s)} \, G(p,s) \,,
\label{SV}
\end{eqnarray}
where $\Phi$ and $G$ are as given below 
\begin{eqnarray}
\Phi(p,s) &\equiv& 
          is \left( p_\parallel^2 - {\tan (e{\cal B}s) \over e{\cal B}s} \, p\pr^2 - m^2 \right) - \epsilon |s| \,,
\label{Phi} \\
G(p,s) &\equiv& {e^{ie{\cal B}s\sigma\!_z} \over \cos(e{\cal B}s)} 
       \, \left( \rlap/p_\parallel + \frac{e^{-ie{\cal B}s\sigma_z}}
{\cos(e{\cal B}s)}\rlap/ p\pr + m \right) \nonumber \\ 
       &=& \Big[ \big( 1 + i\sigma_z \tan (e{\cal B}s) \big)
(\rlap/p_\parallel + m ) +\sec^2(e{\cal B}s) \rlap/ p\pr \Big] \,, 
\label{C}
\end{eqnarray}
where
\begin{eqnarray}
\sigma_z = i\gamma_1 \gamma_2 = - \gamma_0 \gamma_3 \gamma_5 \,,
\label{sigz}
\end{eqnarray}
and we have used,
\begin{eqnarray}
e^{ie{\cal B}s\sigma_z} = \cos( e{\cal B}s) + i\sigma_z \sin(e{\cal B}s) \,.
\end{eqnarray}
To make the expressions transparent we specify our convention in the
following way,
\begin{eqnarray}
\rlap/ p_\parallel &=& \gamma_0 p^0 + \gamma_3 p^3 \nonumber \\                      
\rlap/p\pr &=& \gamma_1 p^1 + \gamma_2 p^2 \nonumber \\
p_\parallel^2 &=& p_0^2 - p_3^2 \nonumber \\
p\pr^2 &=& p_1^2 + p_2^2 \nonumber.
\end{eqnarray}
Of course in the range of integration indicated in Eq.~(\ref{SV}) $s$
is never negative and hence $|s|$ equals $s$.
In the presence of a background medium, the above
propagator is now modified to~\cite{elmf}:
\begin{eqnarray}
iS(p) = iS_B^V(p) + S_B^\eta(p) \,,
\label{fullprop}
\end{eqnarray}
where
\begin{eqnarray}
S_B^\eta(p) \equiv - \eta_F(p) \left[ iS_B^V(p) - i\overline S_B^V(p) \right] \,,
\end{eqnarray}
and 
\begin{eqnarray}
\overline S_B^V(p) \equiv \gamma_0 S^{V \dagger}_B(p) \gamma_0 \,,
\label{Sbar}
\end{eqnarray}
for a fermion propagator, such that
\begin{eqnarray}
S_B^\eta(p) = - \eta_F(p) \int_{-\infty}^\infty ds\; e^{\Phi(p,s)} G(p,s) \,.
\label{Seta}
\end{eqnarray}
Here $\eta_F(p)$ contains the distribution function for the fermions and the anti-fermions:
\begin{eqnarray}
\eta_F(p) &=& \Theta(p\cdot u) f_F(p,\mu,\beta) \nonumber \\
&+& \Theta(-p\cdot u) f_F(-p,-\mu,\beta) \, ,
\label{eta}
\end{eqnarray}
 $f_F$ denotes the Fermi-Dirac distribution function:
\begin{eqnarray}
f_F(p,\mu,\beta) = {1\over e^{\beta(p\cdot u - \mu)} + 1} \,,
\label{distrib}
\end{eqnarray}
and $\Theta$ is the step function given by:
\begin{eqnarray}
\Theta(x) &=& 1, \; \mbox{for $x > 0$} \,, \nonumber \\
&=& 0, \; \mbox{for $x < 0$} \,. \nonumber                
\end{eqnarray}
Here the four velocity of the medium is $u$, in the rest frame it
looks like $u^{\mu}=(1,0,0,0)$. 
%
\subsection{The expression for $\Pi^5_{\mu \nu}$ in thermal medium and
in the presence of a background uniform magnetic field}
The axial polarisation tensor $\Pi^5_{\mu \nu}$ is expressed as
\begin{eqnarray}
i\Pi^5_{\mu \nu}=(-i e)^2(-1) \int {{d^4 p}\over {(2\pi)^4}}\mbox{Tr}\left
[ \gamma_\mu \gamma_5 iS(p) \gamma_\nu iS(p')\right].
\label{pi5}
\end{eqnarray}
Leaving out the vacuum contribution (the contribution devoid of any
thermal effects) and the  contributions with two thermal factors, 
we are left with
\begin{eqnarray}
i\Pi^5_{\mu\nu}(k)&=&(-i e)^2(-1) \int {{d^4 p}\over {(2\pi)^4}}\mbox{Tr}\left
[ \gamma_\mu \gamma_5 S^\eta_B(p) \gamma_\nu
iS^V_B(p')\right.\nonumber\\
&+&\left. \gamma_\mu
\gamma_5 iS^V_B(p) \gamma_\nu S^\eta_B(p')\right].
\label{pi-ini}
\end{eqnarray}
The vacuum part has already been done in {\cite{hari}} and the thermal part
is related with pure absorption effects in the medium, which we are leaving
out for the time being.

Using the form of the fermion propagator in a magnetic field in
presence of a thermal medium, as given by expressions(\ref{SV}) and (\ref{Seta})
we get
\begin{eqnarray}
i\Pi^5_{\mu\nu}(k)&=&-(-i e)^2(-1) \int {{d^4 p}\over {(2\pi)^4}}
\int_{-\infty}^\infty ds\, e^{\Phi(p,s)}\nonumber\\
& &\int_0^\infty
ds'e^{\Phi(p',s')}\big[ \mbox{Tr}\left[\gamma_\mu\gamma_5 G(p,s)
\gamma_\nu G(p',s')\right]\eta_F(p)\big. \nonumber\\
& &\big.\hskip 1cm + \mbox{Tr}\left[\gamma_\mu
\gamma_5 G(-p',s') \gamma_\nu G(-p,s)\right]\eta_F(-p)
 \big]\nonumber\\
 &=& -(-i e)^2(-1) \int {{d^4 p}\over {(2\pi)^4}}
\int_{-\infty}^\infty ds\, e^{\Phi(p,s)}\nonumber\\
& & \times  \int_0^\infty
ds'\,e^{\Phi(p',s')}\mbox{R}_{\mu\nu}(p,p',s,s')
\label{compl}
\end{eqnarray}
where $\mbox{R}_{\mu\nu}(p,p',s,s')$ contains the trace part. 
\subsection{$\mbox{R}_{\mu\nu}$ to even and odd orders in magnetic field}
We calculate $\mbox{R}_{\mu\nu}(p,p',s,s')$ to even and odd orders in the
external magnetic field and call them $\mbox{R}^{(e)}_{\mu\nu}$
and $\mbox{R}^{(o)}_{\mu\nu}$. The reason for doing this is 
that the two contributions have different properties as far as their
dependence on medium is concerned, a topic which will be discussed
in the concluding section. Calculating the traces we obtain,
\begin{eqnarray}
& &\mbox{R}^{(e)}_{\mu\nu}=
4i\eta_{-}(p)\left[\varepsilon_{\mu \nu \alpha\para \beta\para}
p^{\alpha\para} p'^{\beta\para}(1 + \tan(e{\cal B}s)\tan(e{\cal B}s'))\right.\nonumber\\
&+&\left. \varepsilon_{\mu \nu \alpha\para
\beta\pr} p^{\alpha\para} p'^{\beta\pr} \sec^2 (e{\cal
B}s')
+\varepsilon_{\mu \nu \alpha\pr \beta\para} p^{\alpha\pr} 
p'^{\beta\para} \sec^2 (e{\cal B}s)\right.\nonumber\\
&+&\left. \varepsilon_{\mu \nu \alpha\pr 
\beta\pr} p^{\alpha\pr} p'^{\beta\pr}\sec^2 (e{\cal
B}s)\sec^2 (e{\cal B}s')\right]
\label{reven}
\end{eqnarray}
and
\begin{eqnarray}
& &\mbox{R}^{(o)}_{\mu\nu}=4i\eta_{+}(p)\left[ m^2\varepsilon_{\mu \nu
1 2}(\tan(e{\cal B}s) + \tan(e{\cal B}s'))\right.\nonumber\\
&+&\left.\left\{(g_{\mu \alpha\para} p^{\widetilde{\alpha\para}}
p'_{\nu\para} - g_{\mu \nu} p'_{\alpha\para}
p^{\widetilde{\alpha\para}} +g_{\nu \alpha\para} p^{\widetilde{\alpha\para}}
p'_{\mu\para} )\right.\right.\nonumber\\
&+&\left.\left.(g_{\mu \alpha\para} p^{\widetilde{\alpha\para}}
p'_{\nu\pr} + g_{\nu \alpha\para} p^{\widetilde{\alpha\para}}
p'_{\mu\pr}) \sec^2(e{\cal B}s')\right\} \tan(e{\cal
B}s)\right.\nonumber\\
&+& \left.\left\{(g_{\mu \alpha\para} p'^{\widetilde{\alpha\para}}
p_{\nu\para} - g_{\mu \nu} p_{\alpha\para}
p'^{\widetilde{\alpha\para}} +g_{\nu \alpha\para} 
p'^{\widetilde{\alpha\para}}
p_{\mu\para} )\right.\right.\nonumber\\
&+&\left.\left.(g_{\mu \alpha\para} p'^{\widetilde{\alpha\para}}
p_{\nu\pr} + g_{\nu \alpha\para} p'^{\widetilde{\alpha\para}}
p_{\mu\pr}) \sec^2(e{\cal B}s)\right\} \tan(e{\cal
B}s') \right]. 
\label{rodd}
\end{eqnarray}
Here
\begin{eqnarray}
\eta_+(p)&=&\eta_F(p) + \eta_F(-p) \label{etaplus}\\
\eta_-(p)&=&\eta_F(p) - \eta_F(-p)
\label{etaminus}
\end{eqnarray}
which contain the information about the distribution functions.
Also it should be noted that, in our convention
\begin{eqnarray}
a_{\mu} b^{{\widetilde \mu}\para}=a_0 b^3 + a_3 b^0.\nonumber
\end{eqnarray}

If we concentrate on the rest frame of the medium, then
$p\cdot u=p_0$. Thus the distribution function does not depend on the
spatial components of $p$. In this case we can write the expressions of
$\mbox{R}^{(e)}_{\mu \nu}$ and $\mbox{R}^{(o)}_{\mu \nu}$ using the
relations derived 
earlier {\cite{frd1}} inside the integral sign, as
\begin{eqnarray}
p^{\beta\pr}&\stackrel{\circ}{=}&-{\tan(e{\cal B}s')\over {\tan(e{\cal B}s) +
\tan(e{\cal B}s')}}\,k^{\beta\pr}\label{pperpint}\\
p'^{\beta\pr}&\stackrel{\circ}{=}&{\tan(e{\cal B}s)\over {\tan(e{\cal B}s) + \tan(e{\cal
B}s')}}\,k^{\beta\pr}\label{primeperpint}\\
p^2\pr&\stackrel{\circ}{=}&{1\over {\tan(e{\cal B}s) + \tan(e{\cal B}s')}}\left[-ie{\cal
B}\right.\nonumber\\
& &\hskip 1.5cm \left. + {\tan(e{\cal B}s')^2\over
{\tan(e{\cal B}s) + \tan(e{\cal B}s')}}\, k^2\pr\right]\label{psq}\\
p'^2\pr&\stackrel{\circ}{=}&{1\over {\tan(e{\cal B}s) + \tan(e{\cal B}s')}}\left[-ie{\cal
B}\right.\nonumber\\
& &\hskip 1.5cm \left. + {\tan(e{\cal B}s')^2\over
{\tan(e{\cal B}s) + \tan(e{\cal B}s)}}\, k^2\pr\right]\label{p'sq}\\
m^2 &\stackrel{\circ}{=}&\left(i{d\over ds} + (p^2\para -\sec^2(e{\cal B}s)
p^2\pr)\right)
\end{eqnarray}
 and get
\begin{eqnarray}
\mbox{R}^{(e)}_{\mu \nu}&\stackrel{\circ}{=}&
4i\eta_{-}(p_0)\left[\varepsilon_{\mu \nu \alpha\para \beta\para}
p^{\alpha\para} p'^{\beta\para}(1 + \tan(e{\cal B}s)\tan(e{\cal B}s'))\right.\nonumber\\
&+&\left. \varepsilon_{\mu \nu \alpha\para
\beta\pr} p^{\alpha\para} p'^{\beta\pr} \sec^2 (e{\cal
B}s')\right.\nonumber\\
&+&\left.\varepsilon_{\mu \nu \alpha\pr \beta\para} p^{\alpha\pr} 
p'^{\beta\para} \sec^2 (e{\cal B}s)\right]
\label{reven1}
\end{eqnarray}
and 
\begin{eqnarray}
\mbox{R}^{(o)}_{\mu \nu}&\stackrel{\circ}{=}&
4i\eta_+(p_0)\left[-\varepsilon_{\mu \nu 1 2} 
\left\{ \frac{\sec^2(e{\cal B}s)\tan^2(e{\cal B}s')}{\tan(e{\cal B}s)
+ \tan(e{\cal B}s')}
k_{\pr}^2 \right.\right.\nonumber\\
&+& \left.\left. (k\cdot p)\para (\tan(e{\cal B}s) +
\tan(e{\cal B}s'))\right\}\right.\nonumber\\ 
&+&\left. 2\varepsilon_{\mu 1 2 \alpha\para}(p'_{\nu\para}
p^{\alpha\para}\tan(e{\cal B}s) +
p_{\nu\para}p'^{\alpha\para}\tan(e{\cal B}s'))\right.\nonumber\\
&+&\left. g_{\mu\alpha\para} k_{\nu\pr}\left\{p^{\widetilde
\alpha\para}(\tan(e{\cal B}s)
 - \tan(e{\cal B}s'))\right.\right.\nonumber\\
&-&\left.\left. k^{\widetilde \alpha\para}\,
{\sec^2(e{\cal B}s)\tan^2(e{\cal B}s')\over{\tan(e{\cal B}s) +
\tan(e{\cal B}s')}}\right\}\right.\nonumber\\
&+&\left.\{g_{\mu\nu}(p\cdot \widetilde k)\para + g_{\nu \alpha\para}
p^{\widetilde \alpha\para} k_{\mu\pr}\} \right.\nonumber\\
&\times& \left.(\tan(e{\cal B}s) - \tan(e{\cal B}s'))\right.\nonumber\\
&+&\left. g_{\nu \alpha\para}
k^{\widetilde\alpha\para}p_{\mu\pr}\sec^2(e{\cal B}s)\tan(e{\cal B}s')\right].
\label{crmunu}
\end{eqnarray}
The $\stackrel{\circ}{=}$ symbol signifies that the above relations are
not proper equations, the equality holds only inside the momentum
integrals in Eq.(\ref{compl}). 
\section{Gauge Invarience}
\label{gaugeinvariance}
\subsection{Gauge invarience for $\Pi^5_{\mu \nu}$ to even
orders in the external field}
The  axial polarisation tensor even in the external field is given by
\begin{eqnarray}
\Pi^{5(e)}_{\mu \nu}&=& -(-i e)^2(-1) \int {{d^4 p}\over {(2\pi)^4}}
\int_{-\infty}^\infty ds\, e^{\Phi(p,s)}\nonumber\\
& & \times  \int_0^\infty
ds'\,e^{\Phi(p',s')}\mbox{R}^{(e)}_{\mu\nu}(p,p',s,s').
\label{a1}
\end{eqnarray}
Using  Eq.(\ref{reven1}) in the rest frame of the medium, we have
\begin{eqnarray}
& &\mbox{R}^{(e)}_{\mu \nu}\stackrel{\circ}{=}
4i\eta_{-}(p_0)\left[\varepsilon_{\mu \nu \alpha\para \beta\para}
p^{\alpha\para} p'^{\beta\para}(1 + \tan(e{\cal B}s)\tan(e{\cal B}s'))\right.\nonumber\\
&+&\left. \varepsilon_{\mu \nu \alpha\para
\beta\pr} p^{\alpha\para} p'^{\beta\pr} \sec^2 (e{\cal
B}s')
+\varepsilon_{\mu \nu \alpha\pr \beta\para} p^{\alpha\pr} 
p'^{\beta\para} \sec^2 (e{\cal B}s)\right].\nonumber\\
\label{a2}
\end{eqnarray}
Noting that it is possible to write, 
\begin{eqnarray}
q^{\alpha}p_{\alpha} = q^{\alpha\para}p_{\alpha\para} + 
q^{\alpha\pr}p_{\alpha\pr}\nonumber
\end{eqnarray}
 Eq.(\ref{a2}) can be written as,  
\begin{eqnarray}
& &\mbox{R}^{(e)}_{\mu \nu}\stackrel{\circ}{=}
4i\eta_{-}(p_0)\left[(\varepsilon_{\mu \nu \alpha \beta}
p^{\alpha} p'^{\beta} - \varepsilon_{\mu \nu \alpha \beta\pr}
p^{\alpha} p'^{\beta\pr}\right.\nonumber\\
&-&\left.\varepsilon_{\mu \nu \alpha\pr \beta}
p^{\alpha\pr} p'^{\beta})(1 + \tan(e{\cal B}s)\tan(e{\cal B}s'))\right.\nonumber\\
&+&\left. \varepsilon_{\mu \nu \alpha
\beta\pr} p^{\alpha} p'^{\beta\pr} \sec^2 (e{\cal
B}s')
+\varepsilon_{\mu \nu \alpha\pr \beta} p^{\alpha\pr} 
p'^{\beta} \sec^2 (e{\cal B}s)\right].\nonumber\\
\label{a3}
\end{eqnarray}
Here throughout we have omitted terms such as $\varepsilon_{\mu \nu
\alpha\pr \beta\pr} p^{\alpha\pr}  
p'^{\beta\pr}$, since  by the application of Eq.(\ref{pperpint}) we
have
\begin{eqnarray}
\varepsilon_{\mu \nu \alpha\pr \beta\pr} p^{\alpha\pr} 
p'^{\beta\pr} &=&\varepsilon_{\mu \nu \alpha\pr \beta\pr}
p^{\alpha\pr} p^{\beta\pr} + \varepsilon_{\mu \nu \alpha\pr
\beta\pr} p^{\alpha\pr}  
k^{\beta\pr} \nonumber\\
&\stackrel{\circ}{=}& - \frac{\tan(e{\cal B}s')}{\tan(e{\cal B}s')+\tan(e{\cal B}s')}
\varepsilon_{\mu \nu \alpha\pr 
\beta\pr} k^{\alpha\pr}  
k^{\beta\pr}\nonumber
\end{eqnarray}
which is zero.

After rearranging the terms appearing in Eq.(\ref{a3}), and by the
application of Eqs.(\ref{pperpint}) and (\ref{primeperpint}) we
arrive at the expression
\begin{eqnarray}
& &\mbox{R}^{(e)}_{\mu \nu}\stackrel{\circ}{=}4i\eta_{-}(p_0)\Bigg[\varepsilon_{\mu \nu
\alpha \beta} 
p^{\alpha} k^{\beta}(1 + \tan(e{\cal B}s)\tan(e{\cal B}s'))\nonumber\\
&+& \varepsilon_{\mu \nu \alpha
\beta\pr} k^{\alpha} k^{\beta\pr} 
\tan(e{\cal B}s) \tan(e{\cal B}s') \frac{\tan(e{\cal B}s)-\tan(e{\cal B}s')}{\tan(e{\cal B}s) + \tan(e{\cal B}s')}\Bigg].\nonumber\\
\label{evenpart}
\end{eqnarray}
Because of the presence of terms like $\varepsilon_{\mu \nu
\alpha \beta} 
 k^{\beta}$ and $ \varepsilon_{\mu \nu \alpha
\beta\pr} k^{\alpha} $ if we contract $\mbox{R}^{(e)}_{\mu \nu}$  by
$k^\nu$, it vanishes.
\subsection{Gauge invarience for $\Pi^5_{\mu \nu}$ to odd
orders in the external field}
The  axial polarisation tensor odd in the external field is given by
\begin{eqnarray}
\Pi^{5(o)}_{\mu \nu}&=& -(-i e)^2(-1) \int {{d^4 p}\over {(2\pi)^4}}
\int_{-\infty}^\infty ds\, e^{\Phi(p,s)}\nonumber\\
& & \times  \int_0^\infty
ds'\,e^{\Phi(p',s')}\mbox{R}^{(o)}_{\mu\nu}(p,p',s,s')
\end{eqnarray}
where $\mbox{R}^{(o)}_{\mu\nu}(p,p',s,s')$ is given by Eq.(\ref{crmunu}).
The general gauge invarience condition in this case
\begin{eqnarray}
k^{\nu} \Pi^{5(o)}_{\mu \nu} &=& 0 
\end{eqnarray}
can always be written down in terms of the following two equations,
\begin{eqnarray}
k^{\nu} \Pi^{5(o)}_{\mu\para \nu} &=& 0 \label{gipara}\\
k^{\nu} \Pi^{5(o)}_{\mu\pr \nu} &=& 0
\label{gipr}
\end{eqnarray}
where $ \Pi^{5(o)}_{\mu\para \nu}$  is that part of $
\Pi^{5(o)}_{\mu \nu}$ where the index $\mu$ can take the values $0$ and 
$3$ only. Similarly 
 $ \Pi^{5(o)}_{\mu\pr \nu}$ stands for the part of $
\Pi^{5(o)}_{\mu \nu}$ where  $\mu$  can take the values $1$ and $2$
only.  $ \Pi^{5(o)}_{\mu\para \nu}$ contains $\mbox{R}^{(o)}_{\mu\para
\nu}(p,p',s,s')$ which from Eq.(\ref{crmunu}) is as follows,
\begin{eqnarray}
\mbox{R}^{(o)}_{\mu\para \nu}&\stackrel{\circ}{=}&
4i\eta_+(p_0)
\left[
-\varepsilon_{\mu\para \nu 1 2} 
\left\{ \frac{\sec^2(e{\cal B}s)\tan^2(e{\cal B}s')}{\tan(e{\cal B}s)
+ \tan(e{\cal B}s')} 
k_{\pr}^2 \right.\right.\nonumber\\
&+& \left.\left. (k\cdot p)\para (\tan(e{\cal B}s) +
\tan(e{\cal B}s'))\right\}\right.\nonumber\\ 
&+&\left. 2\varepsilon_{\mu\para 1 2 \alpha\para}\,(p'_{\nu\para}
p^{\alpha\para}\tan(e{\cal B}s) +
p_{\nu\para}p'^{\alpha\para}\tan(e{\cal B}s'))\right.\nonumber\\
&+&\left. g_{\mu\para \alpha\para} k_{\nu\pr}\left\{p^{\widetilde
\alpha\para}(\tan(e{\cal B}s)
 - \tan(e{\cal B}s'))\right.\right.\nonumber\\
&-&\left.\left. k^{\widetilde \alpha\para}\,
{\sec^2(e{\cal B}s)\tan^2(e{\cal B}s')\over{\tan(e{\cal B}s) +
\tan(e{\cal B}s')}}\right\}\right.\nonumber\\
&+&\left. g_{\mu\para\nu}(p\cdot \widetilde k)\para(\tan(e{\cal B}s) -
\tan(e{\cal B} s'))
\right]
\label{para}
\end{eqnarray}
and $ \Pi^{5(o)}_{\mu\pr \nu}$ contains
$\mbox{R}^{(o)}_{\mu\pr \nu}(p,p',s,s')$  which is 
\begin{eqnarray}
{\mbox R}^{(o)}_{\mu\pr
\nu}&\stackrel{\circ}{=}&4i\eta_+(p_0)\left[\{g_{\mu\pr \nu}(p\cdot
\widetilde k)\para + g_{\nu \alpha\para} p^{\widetilde \alpha\para}
k_{\mu\pr}\}\right.\nonumber\\
& &\hskip 1cm \times  \left.(\tan(e{\cal B}s) - \tan(e{\cal B}s'))\right.\nonumber\\ 
&+& \left.g_{\nu \alpha\para}
k^{\widetilde\alpha\para}p_{\mu\pr}\sec^2(e{\cal B}s)\tan(e{\cal B}s')\right]. 
\label{perp}
\end{eqnarray}
Eqs.(\ref{gipara}),(\ref{gipr})  implies one should have the
following relations satisfied,
\begin{eqnarray}
k^{\nu} \int {{d^4 p}\over {(2\pi)^4}} 
\int_{-\infty}^\infty ds\, e^{\Phi(p,s)} \int_0^\infty
ds'\,e^{\Phi(p',s')} \, 
{\mbox R}^{(o)}_{\mu\pr \nu}=0
\nonumber \\
\label{gipr1}
\end{eqnarray}
and
\begin{eqnarray}
k^{\nu}\int {{d^4 p}\over {(2\pi)^4}} 
\int_{-\infty}^\infty ds\, e^{\Phi(p,s)} \int_0^\infty
ds'\,e^{\Phi(p',s')}\, {\mbox R}^{(o)}_{\mu\para \nu}=0.
\label{gipara1}
\end{eqnarray}
Out of the two above equations,  Eq.(\ref{gipr1}) can be verified 
easily since
\begin{eqnarray}
k^{\nu}{\mbox R}_{\mu\pr \nu}=0.
\end{eqnarray}

Now we look at Eq.(\ref{gipara1}). 
We explicitly consider the case $\mu_{\parallel}=3$ (the
$\mu_{\parallel}=0$ case lead to similar result). For
$\mu_{\parallel}=3$
\begin{eqnarray}
k^{\nu}{\mbox R}^{(o)}_{3 \nu}&\stackrel{\circ}{=}&-p_0\left[(p'^{\,2}\para -
p^2\para)(\tan(e{\cal B}s) + \tan(e{\cal B}s')) \right.\nonumber\\
&-&\left. k^2\pr(\tan(e{\cal B}s) - \tan(e{\cal B}s'))\right](4i \eta_+(p_0)).\nonumber\\
\label{a_1}
\end{eqnarray}

Apart from the small convergence factors, 
\begin{eqnarray}
{i \over e{\cal B}} && \left(\Phi(p,s) + \Phi(p',s')\right) \nonumber \\
&=& \left( p_\parallel^{\prime2} + p_\parallel^2 - 2m^2 \right) \xi 
- \left(p_\parallel^{\prime2} - p_\parallel^2 \right) \zeta \nonumber \\
&& - p\pr^{\prime2} \tan (\xi-\zeta) - p\pr^2 \tan (\xi+\zeta) \,,
\end{eqnarray}
\label{newa}
where we have defined the parameters
\begin{eqnarray}
\xi &=& \frac12 e{\cal B}(s+s') \,, \nonumber\\*
\zeta &=& \frac12 e{\cal B}(s-s') \,.
\label{xizeta}
\end{eqnarray}
From the last two equations we can write
\begin{eqnarray}
&& {ie{\cal B}} \; {d\over d\zeta} e^{\Phi(p,s) + \Phi(p',s')} \nonumber \\
&& = e^{\Phi(p,s) + \Phi(p',s')} \nonumber \\
&& \times \, \left(p_\parallel^{\prime2} - p_\parallel^2 - p\pr^{\prime2} \sec^2 (\xi-\zeta) + p\pr^2 \sec^2 (\xi+\zeta) \right) \,
\label{C1par}
\end{eqnarray}
which implies
\begin{eqnarray}
p'^{\,2}\para - p^2\para &=&
\nonumber \\
 ie{\cal B}{d\over{d\xi}} &+& \left[ p'^{\,2}\pr
\sec^2(e{\cal B}s') - p^2\pr \sec^2(e{\cal B}s)\right].       
\label{a_parsqrdiff}
\end{eqnarray}
The equation above is valid in the sense that both sides of it
actually acts upon $e^{\widetilde \Phi(p,s,p',s')}$, where
\begin{eqnarray}
\widetilde \Phi(p,p',s,s') = \Phi(p,s) + \Phi(p',s').
\label{a_Phi}
\end{eqnarray}
From Eqs.(\ref{a_1}) and (\ref{a_parsqrdiff}) we have
\begin{eqnarray}
k^{\nu} \, {\mbox R}_{3 \,\nu}\, e^{\widetilde \Phi}
&\stackrel{\circ}{=}&-4i\eta_+(p_0)p_0\left[(p'^{\,2}\pr \sec^2(e{\cal B}s) - p^2\pr
\sec^2(e{\cal B}s))\right.\nonumber\\
& & \hskip 1cm \left.\times (\tan(e{\cal B}s)+\tan(e{\cal B}s'))\right.\nonumber\\
&-&\left. k^2\pr(\tan(e{\cal B}s)-\tan(e{\cal B}s'))\right.\nonumber\\
&+&\left.ie{\cal B}p_0(\tan(e{\cal B}s) + \tan(e{\cal B}s'))\,{d
\over{d \xi}}\right] e^{\widetilde 
\Phi}.\nonumber\\
\label{a_kdotR1}
\end{eqnarray}
Now using the the expressions for $p^2\pr$ and $p'^2\pr$ from
Eqs.(\ref{psq}) and (\ref{p'sq}) 
we can write
\begin{eqnarray}
k^{\nu} \, {\mbox R}_{3 \,\nu}\,e^{\widetilde \Phi}&\stackrel{\circ}{=}&4e{\cal B}\eta_+(p_0)
p_0\left[(\sec^2(e{\cal B}s) - \sec^2(e{\cal B}s'))\right.\nonumber\\
&+&\left. (\tan(e{\cal B}s) + \tan(e{\cal B}s')){d
\over{d \xi}}\right] e^{\widetilde \Phi} .\nonumber\\
\label{a_kdotR2}
\end{eqnarray}
The above equation can also be written as 
\begin{eqnarray}
k^{\nu} \, {\mbox R}_{3 \,\nu}\, e^{\widetilde
\Phi}\stackrel{\circ}{=}4e{\cal B} \eta_+(p_0)p_0 
{d \over{d \xi}} \left[ e^{\widetilde \Phi} (\tan(e{\cal B}s) +
\tan(e{\cal B}s'))\right].
\end{eqnarray}
Transforming to $\xi, \zeta$ variables and using the above
equation we can write the parametric integrations (integrations over
$s$ and $s'$) on the left hand side of Eq.(\ref{gipara1}) as
\begin{eqnarray}
& &\int_{-\infty}^{\infty} ds \int_0^{\infty} ds' k^{\nu} \, {\mbox R}_{3\,
\nu}\, e^{\widetilde \Phi}\nonumber\\
&=&\frac{8\eta_+(p_0)p_0}{e{\cal B}}\int_{-\infty}^{\infty} d\xi
\int_{-\infty}^{\infty} d\zeta \Theta 
(\xi - \zeta) {d\over{d\xi}}{\cal F}(\xi,\zeta)\nonumber\\
\label{odd}
\end{eqnarray}
where
\begin{eqnarray}
{\cal F}(\xi,\zeta)=e^{\widetilde \Phi} (\tan(e{\cal B}s) + \tan(e{\cal B}s')).\nonumber
\end{eqnarray}
The integration over the $\xi$ and $\zeta$ variables in Eq.(\ref{odd}) 
can be represented as,
\begin{eqnarray}
& &\int_{-\infty}^{\infty} d\xi \int_{-\infty}^{\infty} d\zeta \Theta
(\xi - \zeta) {d\over{d\xi}}{\cal
F}(\xi,\zeta)\nonumber\\
&=&\int_{-\infty}^{\infty} d\xi \int_{-\infty}^{\infty}
d\zeta \left[{d\over{d\xi}}\{ \Theta(\xi - \zeta) {\cal F}(\xi,\zeta)\}
- \delta(\xi - \zeta) {\cal F}(\xi,\zeta)\right]\nonumber\\
&=&-\int_{-\infty}^{\infty} d\xi {\cal F}(\xi,\xi)
\label{end1}
\end{eqnarray}
here the second step follows from the first one as the first integrand
containing the $\Theta$ function vanishes at both limits of the
integration. The remaining integral is now only a function of $\xi$
and is even in $p_0$. But in Eq.(\ref{odd}) we have $\eta_+(p_0)p_0$
sitting, which makes the the integrand odd under $p_0$ integration
in the left hand side of Eq.(\ref{gipara1}), as $\eta_+(p_0)$ is an even function in
$p_0$. So the $p_0$ integral as it occurs in the left hand side of equation 
(\ref{gipara1}) vanishes as expected, yeilding the required result shown in
Eq.(\ref{gipara}). 
\section{Effective Charge}
\label{efftcharge}
Now we concentrate on the neutrino effective charge. 
From the onset it is to be made clear that 
we are only calculating the axial contribution to the effective
charge.{\footnote {In a forthcoming publication we will comment on the vector
contribution to the effective charge of the neutrino{\cite{bg}}.}}
 We can now write the full expression of the axial polarisation tensor 
as
\begin{eqnarray}
i\Pi^5_{\mu\nu}(k)&=& -(-i e)^2(-1) \int {{d^4 p}\over {(2\pi)^4}}
\int_{-\infty}^\infty ds\, e^{\Phi(p,s)}\nonumber\\
& & \times  \int_0^\infty
ds'\,e^{\Phi(p',s')}\left[\mbox{R}^{(o)}_{\mu\nu} + \mbox{R}^{(e)}_{\mu\nu}\right]
\label{alord}
\end{eqnarray}
where $\mbox{R}^{(o)}_{\mu\nu}$ and
$\mbox{R}^{(e)}_{\mu\nu}$ are given by
Eqs.(\ref{crmunu}) and (\ref{reven1}) in the rest frame of the
medium.
\subsection{Effective Charge to odd orders in external field}
In the
limit when the external momentum tends to zero only two terms
survive from  $\Pi^{5}_{\mu\nu}(k)$.  Denoting
$\Pi^{5}_{\mu\nu}(k_0=0, {\vec k } \to 0) = \Pi^{5}_{\mu\nu}$,  we obtain 
\begin{eqnarray}
\Pi^5_{\mu 0}&=&\lim_{k_0=0
\vec{k}\rightarrow 0}4 e^2 \int{d^4p\over{(2\pi)^4}} 
\int^{\infty}_{-\infty} ds\, e^{\Phi(p,s)}
\nonumber \\ 
&\times&\int^{\infty}_0 ds'
e^{\Phi(p',s')}(\tan(e{\cal B}s) + \tan(e{\cal B}s')) \nonumber\\
&\times &\eta_+(p_0)\left[2 p^2_0 -  (k\cdot p)\para \right]\varepsilon_{\mu 0 1 2}
\label{pi5k0}
\end{eqnarray}
the other terms turns out to be zero in this limit. The above equation
shows that, except the exponential functions, the integrand is free of
the perpendicular components of momenta. This implies we can integrate
out the perpendicular component of the loop momentum.
Upon performing the  gaussian integration over the perpendicular components
and taking the limit $k_{\pr} \to 0$, we obtain,
\begin{eqnarray}
\Pi^5_{\mu 0}&=&\lim_{k_0=0,\vec{k}\rightarrow 0} \frac{(-4i e^3 B)}{4\pi} \int{d^2 p\para \over{(2\pi)^2}}
\int^{\infty}_{-\infty} ds
\nonumber \\ &&\times
\, e^{is(p^2\para - m^2) - \varepsilon|s|}
\int^{\infty}_0 ds'
e^{is'(p'^{\,2}\para - m^2) - \varepsilon|s'|} \nonumber\\
&\times &\eta_+(p_0)\left[2 p^2_0 -  (k\cdot p)\para \right]\varepsilon_{\mu 0 1 2}.
\label{charge1}
\end{eqnarray}
It is worth noting that the $s$ integral gives
\begin{eqnarray}
\int^{\infty}_{\infty} ds\, e^{is(p^2\para - m^2) - \varepsilon|s|} =
2\pi \delta(p^2\para - m^2)
\label{delta}
\end{eqnarray}
and the $s'$ integral gives
\begin{eqnarray}
\int^{\infty}_0 ds'\, e^{is'(p'^{\,2}\para - m^2) - \varepsilon|s'|} =
{i\over{(p'^{\,2}\para - m^2) + i\varepsilon}}.
\label{divergent}
\end{eqnarray}
Using the above results in Eq.(\ref{charge1}) and using the delta
function constraint, we arrive at,
\begin{eqnarray}
\Pi^5_{\mu 0} &=&
\lim_{k_0=0,\vec{k}\rightarrow 0} 
2(e^3 B) \int{d^2 p\para \over{(2\pi)^2}}
{\delta(p^2\para - m^2)}
\eta_+(p_0)
\nonumber \\ &\times&
\Bigg[{2 p^2_0\over{(k^{\,2}\para +2(p.k)\para)}}
 -  \frac{1}{2}\Bigg]\varepsilon_{\mu 0 1 2}.
\label{charge2}
\end{eqnarray}
In deriving Eq.[\ref{charge2}], pieces proportional to $k^2\para$  in the 
numerator were neglected.
Now if one makes the substitution, $ p'\para \to (p\para + k\para/2) $ and sets
$k_0 =0$ one arrives at,
\begin{eqnarray}
\Pi^5_{\mu 0} 
&=&
-\lim_{k_0=0,\vec{k}\rightarrow 0} 
2(e^3 {\cal B}) \int{d p_3 \over{(2\pi)^2}}
\left(n_+(E'_p)+ n_-(E'_p)\right)
\nonumber \\ &\times&
\Bigg[ \frac{E'_p}{p_3k_3}
 + \frac{1}{2E'_p}\Bigg]\varepsilon_{\mu 0 1 2}.
\label{charge3}                
\end{eqnarray}
Here  $n_{\pm}(E'_p)$ are the functions $f_F(E'_p,-\mu,\beta)$,
and $f_F(E'_p,\mu,\beta)$, as given in Eq.(\ref{distrib}), which are
nothing but the Fermi-Dirac distribution functions of the particles
and the antiparticles in the medium.
The new term $E'_p$  is defined as follows,
\begin{eqnarray}
E'^2_p=[(p_3-k_3/2)]^2 + m^2, \nonumber
\end{eqnarray}
and it can be expanded for small external
momenta in the following way
\begin{eqnarray}
 E'^2_p\simeq p^2_3+m^2 - p_3 k_3 = E^2_p - p_3 k_3\nonumber
\end{eqnarray}
where $E^2_p = p^2_3+m^2$.
Noting, that
\begin{eqnarray}
E'_p= E_p -  {p_3 k_3 \over 2 E_p} + O(k^2_3),
\label{epprime}
\end{eqnarray}
one can use this expansion in Eq.[\ref{charge3}], to arrive at:
\begin{eqnarray}
\Pi^5_{\,\,\mu 0} 
&=&
-\lim_{k_0=0,\vec{k}\rightarrow 0} 
2(e^3 {\cal B}) \int{d p_3 \over{(2\pi)^2}}
\left(n_+(E'_p)+ n_-(E'_p)\right) 
\nonumber \\ &\times&
\Bigg[ \frac{E_p}{p_3 k_3}
 \Bigg]\varepsilon_{\mu 0 1 2}.
\label{charge5}
\end{eqnarray}
The expression for for $\eta_+(E'_p) = n_+(E'_p)+ n_-(E'_p)$ when
expanded in powers of the 
external momentum $k_3$ is given by
\begin{eqnarray}
\eta_+(E'_p) = ( 1 + \frac{1}{2} \frac{\beta p_3 k_3}{E_p}) \eta_+(E_p)
\label{neweta}
\end{eqnarray}
up to first order terms in the external momentum $k_3$.

\subsubsection{Effective Charge For $\mu \ll m$}

In the limit, when $\mu \ll m$ one can use the 
following expansion,
\begin{eqnarray}
\eta_+(E'_p) &=&\Bigg[ n_{+}(E'_p) +n_{-}(E'_P)\Bigg]\nonumber\\
&=& 2 \sum_{n=0}^{\infty} (-1)^n \cosh([n+1]\beta\mu)e^{-(n+1)\beta E_p}
\nonumber \\ &\times&
\left(1 +\frac{\beta p_3 k_3}{2E_p} + O(k_3^2)+ .....\right)
\label{expan}
\end{eqnarray}

Now  using Eq.(\ref{expan}) in Eq.(\ref{charge5}) we get
\begin{eqnarray}
\Pi^5_{\mu 0} &=&
-\varepsilon_{\mu 0 1 2}\lim_{k_0=0, \vec{k}\rightarrow 0} 
(4e^3 {\cal B}) 
\sum_{n=0}^{\infty} (-1)^n\nonumber\\
 &\times& \cosh([n+1]\beta\mu)
\int{d p_3 \over{(2\pi)^2}}e^{-(n+1)\beta E_p}
\nonumber \\
 &\times& \Bigg[ \frac{E_p}{(p_3 k_3)} + \frac{\beta}{2}
 \Bigg].
\label{charge6}
\end{eqnarray}
The first term vanishes by symmetry of the integral, but the second term 
is finite and so we get:
\begin{eqnarray}
\Pi^5_{\mu 0} &=& -\beta \varepsilon_{\mu 0 1 2}
\lim_{k_0=0 {\vec k}\to 0}
\frac{(e^3 {\cal B})}{2\pi^2} 
\sum_{n=0}^{\infty} (-1)^n \cosh([n+1]\beta\mu)
\nonumber
\\ &\times&
\int d p_3  e^{-(n+1)\beta E_p}.
\label{charge7}
\end{eqnarray} 

To perform the momentum integration, use of the following integral transform
turns out to be extremely convenient
\begin{eqnarray}
e^{-\alpha \sqrt{s}} =\frac{\alpha}{2\sqrt{\pi}} 
\int^{\infty}_0 du e^{-us - \frac{\alpha^2}{4u}} u^{-3/2}.  
\end{eqnarray}
Identifying $\sqrt{s}$ with $E_p$ and $[(n+1)\beta]$ as $\alpha$ (since the square
root opens up), one can easily perform the gaussian $p_3$ integration 
without any difficulty. The result is:
\begin{eqnarray}
\Pi^{5}_{\mu 0}&=& -\beta \varepsilon_{\mu 0 1 2}
\frac{(e^3 {\cal B})}{2\pi^2} 
\sum_{n=0}^{\infty} (-1)^n \cosh([n+1]\beta\mu)
\nonumber
\\ &\times&
 (\beta(n+1)/2)\int du e^{- m^2 u - \frac{((n+1)\beta/2)^2}{u}} u^{-2}.
\label{charge8}
\end{eqnarray}
Performing the the u integration
the axial part of the effective charge of neutrino in the limit of
$m > \mu $ turns out to be,
\begin{eqnarray}
e^{\nu_a}_{\rm eff} &=& -  \sqrt{2} g_{A} m \beta G_F
\frac{e^2 {\cal B}}{\pi^2} (1-\lambda) \cos(\theta)\nonumber\\
&\times&\sum_{n=0}^{\infty} (-1)^n \cosh((n+1)\beta\mu)  
 K_{-1}(m \beta(n+1)).
\label{nucharge}
\end{eqnarray}
Here $\theta$ is the angle between the neutrino three momentum and the
background magnetic field. The superscript $\nu_a$ on $e^{\nu_a}_{\rm
eff}$ denotes that we are calculating the axial contribution of the
effective charge. $K_{-1}(m \beta(n+1))$ is the
modified Bessel  function (of the second kind) of order one (for this function $K_{-1}(x) =
K_{1}(x)$) which sharply falls off as we move away from the
origin in the positive direction. 
Although as temperature tends to zero 
Eq.[\ref{nucharge}] seems to blow up because of the
presence of $m \beta$, but
 $K_{-1}(m\beta(n+1))$ would damp it's growth as $e^{-m\beta}$, 
hence the result remains finite.
\subsubsection{Effective Charge For, $\mu \gg m$}
Here we would try to estimate neutrino effective charge when $\mu \gg
m$ and $ \beta \ne \infty $.  We would like to emphasize that the
last condition should be strictly followed, i.e, temperature $ T \ne 0
$.
Using Eqs.(\ref{charge5}) and (\ref{neweta}) we would obtain 
\begin{eqnarray}
\Pi^5_{3 0} = \frac{e^3 {\cal B}}{
2\pi} \beta \int \frac{dp}{2\pi} \eta_+(E_p).
\label{one}
\end{eqnarray}
Neglecting  $m$ in the expression in $E_p$ we would obtain, 
\begin{eqnarray}
\Pi^5_{3 0} = \frac{e^3 {\cal B}}{
2\pi^2} \ln[(1 +  e^{\beta \mu})( 1 +  e^{-\beta \mu})].
\end{eqnarray}
Same can also be written as
\begin{eqnarray}
\Pi^5_{3 0} 
=\frac{e^3 {\cal B}}{\pi^2} \ln \left(2\cosh(\frac{\beta \mu}{2})\right).
\label{ra}
\end{eqnarray}
The expression for the effective charge then turns out to be
\begin{eqnarray}
e^{\nu_a}_{\rm eff} = - \sqrt{2} g_A G_F \frac{e^2 {\cal B}}{\pi^2} \ln
\left(2\cosh(\frac{\beta \mu}{2})\right) (1- \lambda) \cos(\theta)
\end{eqnarray}
where $\lambda$ is the helicity of the neutrino spinors.

\subsection{Effective Charge At Even Order In The External Field
And Coupling With Magnetic Fields}

From the part of $\Pi_{\mu \nu}$ which is even in the external fields
we see from Eq.(\ref{evenpart}) that
\begin{eqnarray}
\mbox{R}^{(e)}_{\mu 0} &=& 4i\eta_-(p_0)
\Bigg[\varepsilon_{\mu 0
\alpha \beta} 
p^{\alpha} k^{\beta}(1 + \tan(e{\cal B}s)\tan(e{\cal B}s'))\nonumber\\
&+&\varepsilon_{\mu 0 \alpha
\beta\pr} k^{\alpha} k^{\beta\pr} 
\tan(e{\cal B}s) \tan(e{\cal B}s') 
 \nonumber \\ &\times& 
\frac{\tan(e{\cal B}s)-\tan(e{\cal B}s')}{\tan(e{\cal B}s) + \tan(e{\cal B}s')}\Bigg].
\label{evchr}
\end{eqnarray}
which shows that $\Pi^5_{\mu\nu}(k)$ to even orders in the external
field will vanish when $k_0 \to 0, \vec{k} \to 0$. This implies that
there will be no contribution to the effective neutrino charge from
the sector which is even in the powers of ${\cal B}$. 

Can the neutrinos which are propagating in a magnetised plasma couple
with the classical magnetic field? The situation is a little bit
subtle here, as the vertex of the neutrinos with the dynamical photons 
do get changed here due to the presence of the magnetic field, but
this change cannot induce any electromagnetic form factor responsible
for coupling of the neutrinos with any magnetic field.
In order to find the effective charge of the neutrinos which couples them with
time independent magnetic field
 one should look for (as given in Eq.(\ref{chargedef})),
the $\Gamma^i$'s, where $i=1,2$. 
A magnetic field in the $z$-direction, is given by a
gauge where $A_1, A_2$ are both non zero, or one of them is
nonzero. So to calculate the charge which is essential for the
neutrino current to couple with a magnetic field,
one has to put the index $\nu = 1,2$ in Eq.(\ref{alord}) and take
the limit $k_0 \to 0, \vec{k} \to 0$ and see which component
of  $\Pi^5_{\mu 1}(k)$
exists in the prementioned momentum limit. For the odd ${\cal B}$ part
we see from Eq.(\ref{crmunu}) that
\begin{eqnarray}
\mbox{R}^{(o)}_{\mu 1} &=& 4i\eta_+(p_0)\left[g_{\mu \alpha\para} k_1
\left\{ p^{\widetilde{\alpha\para}}(\tan(e{\cal B}s) - \tan(e{\cal
B}s'))\right.\right.\nonumber\\
&-&\left.\left. k^{\widetilde{\alpha\para}}
\frac{\sec^2(e{\cal B}s)\tan^2(e{\cal B}s')}{\tan(e{\cal B}s) +
\tan({e\cal B}s')} \right\}\right.\nonumber\\
&+&\left. g_{\mu 1} (p\cdot {\widetilde{k}})(\tan(e{\cal B}s) -
\tan(e{\cal B}s'))\right]  
\label{magch1}
\end{eqnarray}
which goes to zero as the photon momentum tends to
zero. By the same argument it follows that for $\nu = 2$ there is 
vanishing contribution. Thus it shows that there is no
effective magnetic coupling from ${\cal B}$ odd part.

For the ${\cal B}$ even part, as is seen from Eq.(\ref{reven1}), that
only $\mbox{R}^{(e)}_{1 2}$ survives, and is given by 
\begin{eqnarray}
\mbox{R}^{(e)}_{1 2} =  4i\eta_-(p_0)\left[\varepsilon_{1 2 0
3}(p\cdot {\widetilde{k}})( 1 + \tan(e{\cal B}s)\tan(e{\cal B}s')) \right],
\label{magch2}
\end{eqnarray}
which also perishes in the limit when the external momentum goes to
zero. So from this we can say that $\Pi^5_{\mu \nu}$ has no
contribution for any charge of the neutrinos which can couple them
with the
magnetic field.
\section{Conclusion}
\label{concl}
In our analysis we have calculated the contributions to
$\Pi^{5(o)}_{\mu\nu}(k)$ to odd and even orders in the external
constant magnetic field. The main reason for doing so is 
the fact that,
 $\Pi^{5(o)}_{\mu\nu}(k)$ and
$\Pi^{5(e)}_{\mu\nu}(k)$, the axial polarisation tensors to odd and
even powers in $e{\cal B}$, have different dependence on the 
background matter. Pieces proportional to even powers in ${\cal B}$ 
are proportional to $\eta_{-}(p_0)$, an odd function of  the 
chemical potential. On the
other hand pieces proportional to odd powers in ${\cal B}$ depend on 
$\eta_{+}(p_0)$, and are even in $\mu$ and as a result 
it survives in the limit 
$\mu \to 0$. As has already been noted, this is a direct 
consequence on the charge and parity symmetries of the underlying
theory.

In a background magnetic field the field dependence of the
form factors, which are usually scalars,  can 
be of the following form:
\begin{eqnarray}
k^{\mu}F_{\mu\nu}F^{\nu\lambda}k^{\lambda} 
\mbox{~~~and~~~}
F_{\mu\nu}F^{\mu\nu}.  
\end{eqnarray}
These forms dosen't exhaust all the possibilities, but whatever they
are they must contain an even number of $F$'s and $k$'s  
and hence they will be even functions of ${\cal B}$. 

Of all  possible
tensorial structures  for the axial polarisation tensor in 
a magnetised plasma, there exists one 
which 
satisfies the current conservation condition in the $\nu$ vertex.
That is given by,
\begin{eqnarray}
\phi_{\mu\nu}=
\epsilon_{\mu\alpha\lambda\sigma}F^{\lambda\sigma}u^{\alpha}\left[
u_{\nu} - \frac{(k.u) k_{\nu}}{k^2}\right]. 
\label{ww}
\end{eqnarray}
 Its worth noting that the first term
in  the square bracket in Eq.(\ref{ww}), which is odd in the external field, survives
 in the zero external momentum limit in the rest frame of the
medium.
The tensorial structures which
are explicitly even in powers of $\cal
B$ do have  
$k$'s also, and so they vanish in the limit when the external momentum 
goes to zero. We have earlier noted that the form factors which exist
in the rest frame of the medium and in the zero momentum limit are
even in powers of the external field.This tells us directly that the axial polarisation tensor
must be odd in the external field in the zero external momentum limit,
a result which we have verified in this work.
 However  it
should be noted that the other form factors contribute  in other
situations like  neutrino Cherenkov radiation, which is not discussed
in this work.

In this work we have elucidated upon the physical significance of the
axial polarisation tensor in various neutrino mediated processes in a
magnetised medium, and explicitly written down its form in a gauge
invariant way. It has been shown that the
part of $\Pi^5_{\mu \nu}$ even in $\cal B$ 
doesn't contribute to the effective electric charge. However it does contribute
to physical proceses e.g., neutrino Cherenkov radiation or  neutrino decay 
in a medium. It is worth noting that  in the low density high temperature
limit, the magnitude of $e^{\nu_a}_{\rm eff}$ can become greater than
the effective charge  of the neutrino in ordinary 
medium provided $e{\cal B}$ is large enough. On the other hand in the
high density limit $e^{\nu_a}_{\rm eff}$ can dominate over the effective 
charge of the neutrino as found in 
an unmagnetized medium, provided the temperature is low enough. However
in standard astrophysical objects, e.g., core of type II Supernova
 temperature is of the order of 30- 60 MeV with Fermi momentum around
300 MeV, for red giants the same are 10 keV and 400 keV, for young
white dwarves temperature is around 0.1 - 1 keV and Fermi momentum
500 keV. In these systems one can have relatively large induced neutrino charge, provided
the field strength is large enough. The effective neutrino photon
coupling in a magnetised medium can also shed 
some light in understanding the observed gamma ray bursts or gamma ray 
repeaters observed in nature. 
\section*{Acknowledgment}
We would like to thank Prof. Palash B. Pal and Prof. Parthasarathi Majumdar
for helpful discussions comments and suggestions.

\end{document}